\documentclass{article}
\usepackage{waspaa21,amsmath,url,times,booktabs,tabularx}
\usepackage{array,multirow}
\usepackage{arydshln}
\usepackage{multicol}
\usepackage{footnote}
\usepackage{float}
\usepackage{cite}
\usepackage{enumitem}

\usepackage[pdftex]{graphicx}

\newcolumntype{C}[1]{>{\centering\arraybackslash}p{#1}}
\newcolumntype{L}[1]{>{\raggedright\arraybackslash}p{#1}}
\newcolumntype{R}[1]{>{\raggedleft\arraybackslash}p{#1}}

\title{MIMII DUE: Sound Dataset for Malfunctioning Industrial Machine Investigation and Inspection with 
Domain Shifts due to Changes in Operational and Environmental Conditions}

\name{
      Ryo Tanabe,
      Harsh Purohit,
      Kota Dohi, 
      Takashi Endo,
}
\secondlinename{
      Yuki Nikaido,
      Toshiki Nakamura,
      and Yohei Kawaguchi
}
\address{ Research and Development Group, Hitachi, Ltd.\\
1-280, Higashi-koigakubo, Kokubunji, Tokyo 185-8601, Japan \\
\{ryo.tanabe.rw, yohei.kawaguchi.xk\}@hitachi.com}

\begin{document}

\ninept
\maketitle

\begin{sloppy}

\begin{abstract}
In this paper, we introduce MIMII DUE, a new dataset for malfunctioning industrial machine investigation and inspection with domain shifts due to changes in operational and environmental conditions. 
Conventional methods for anomalous sound detection face practical challenges because the distribution of features changes between the training and operational phases (called domain shift) due to various real-world factors. 
To check the robustness against domain shifts, we need a dataset that actually includes domain shifts, but such a dataset does not exist so far. 
The new dataset we created consists of the normal and abnormal operating sounds of five different types of industrial machines under two different operational/environmental conditions
(source domain and target domain) independent of normal/abnormal, with domain shifts occurring between the two domains. 
Experimental results showed significant performance differences between the source and target domains, indicating that the dataset contains the domain shifts.
These findings demonstrate that the dataset will be helpful for checking the robustness against domain shifts.
\end{abstract}


\section{Introduction}
\label{sec:intro}

Anomalous sound detection (ASD)~\cite{koizumi2017neyman, kawaguchi2017how, koizumi2019neyman, kawaguchi2019anomaly, koizumi2019batch, suefusa2020anomalous, purohit2020deep} is defined as identifying whether the sound emitted from a machine is normal or anomalous.
Automatically detecting machine malfunctions is an essential technology for the fourth industrial revolution, such as factory automation using artificial intelligence (AI), 
and quickly detecting machine abnormalities by observing machine sounds will be helpful for machine condition-based maintenance (CBM).

One of the significant challenges in putting ASD to practical use is how to detect unknown abnormal sounds in a situation where only normal sounds are available as training data (called ``unsupervised'' ASD).
In real-world factories, actual anomalous sounds rarely occur and are highly diverse.
Therefore, it is practically impossible to collect all possible abnormal sound patterns as training data, and it is necessary for unseen abnormal sounds to be detected. 
There are several datasets of machine operating sounds~\cite{koizumi2019toyadmos, purohit2019mimii, grollmisch2019sounding}, 
among which ToyADMOS~\cite{koizumi2019toyadmos} and MIMII Dataset~\cite{purohit2019mimii} are suitable for unsupervised ASD performance evaluation.
Both datasets contain not only many normal sound clips for training and testing but also machine fault sounds for testing.
The two datasets were also used in Task 2 of the IEEE AASP Challenge on Detection and Classification of Acoustic Scenes and Events (DCASE 2020 Challenge)~\cite{koizumi2020dcase},
and experiments on the two datasets have led to the development of various methods for unsupervised ASD~\cite{giri2020self, primus2020anomalous, kapka2020id, inoue2020detection}.

The other significant challenge is how to perform ASD even under domain-shifted conditions, i.e., when the acoustic characteristics are different between training and monitoring (and, of course, testing before monitoring).
We have to ensure that normal sounds are not incorrectly judged as anomalous due to changes within normal conditions. 
Real-world cases often involve different machine operating conditions between the training and monitoring phases. 
A typical example of this is when the motor speed continuously varies in a conveyor transporting products on a production line based on the production volume in response to product demand. 
Since there is infinite variation in rotation speed, the sound will also change with infinite variation. 
Due to the seasonal demand for certain products, a limited period of training data limits the motor speed during that period (e.g., 200--300 rpm for autumn), 
which also limits the distribution of the training data.
However, in the monitoring phase, the ASD system must continue to monitor the conveyor through all seasons, so it must monitor all possible motor speed conditions, including those that differ from the training data (such as 100--400 rpm). 
In addition to the machine conditions, environmental noise conditions such as signal-to-noise ratio (SNR) and sound characteristics also fluctuate uncontrollably depending on the seasonal demand. 
In such a situation, the normal state's distribution will change.
Neither the ToyADMOS nor the MIMII dataset is designed to evaluate robustness to domain shifts. 
In other words, the conventional datasets do not contain test data recorded under conditions different from those under which the training data were recorded.
Some datasets focus on a kind of domain shift called recording device mismatch~\cite{annamaria2018multi, heittola2020acoustic}, but they are datasets for scene classification, not for anomalous sound detection.

In this paper, we introduce a new dataset we created for checking the robustness of anomalous sound detection against domain shifts.
We call this dataset ``Sound dataset for malfunctioning industrial machine investigation and inspection with domain shifts due to changes in operational and environmental conditions'' (MIMII DUE). 
It consists of the normal and abnormal operating sounds of five different types of industrial machines.
The data for each machine type include six subsets called ``sections'', and each section roughly corresponds to a single product.
Each section consists of data from two domains, i.e., the source domain and the target domain, with different conditions such as operating speed and environmental noise.
The entire dataset contains more than 420,000 seconds of data.
Experimental results showed significant performance differences between the source and target domains, indicating that the dataset contains the domain shifts.
Hence, we expect this dataset will be helpful for checking the robustness against domain shifts.
The dataset is freely available for download at \url{https://zenodo.org/record/4740355} and is a subset of the dataset for Task 2 of the DCASE 2021 Challenge~\cite{kawaguchi2021dcase}.

\setlength{\tabcolsep}{1mm} 
\begin{table*}[t!]
\begin{center}
\caption{Overview of MIMII DUE dataset. 
``Nrm'' and ``Abn'' mean normal and abnormal, respectively.
The test data in sections 03 to 05 is used as evaluation data for DCASE 2021 Challenge Task 2, so the breakdown between normal and abnormal is kept secret until the end of the challenge.
``$c \pm d$'' in SNR  means that the $\gamma$ for each clip was drawn from the uniform distribution over $\left[c - d, c + d\right)$.}
\label{tab:overview}
\footnotesize
\begin{tabular}{ll || rp{2pt}rrr | rp{2pt}rrr | p{7.0cm}}
\hline
	\multicolumn{2}{@{}l ||}{\multirow{4}{*}{ \begin{tabular}{@{\hskip0pt}l@{\hskip0pt}} Machine \\ type/ \\ section \\ index \end{tabular} }} &
	\multicolumn{5}{c |}{Source domain} &
	\multicolumn{5}{c |}{Target domain} &
	\multirow{4}{*}{Description} \\
\cline{3-7} \cline{8-12}
	&
	&
	\multicolumn{4}{c}{\# of 10-second clips} &
	\multirow{3}{*}{SNR [dB]} &
	\multicolumn{4}{c}{\# of 10-second clips} &
	\multirow{3}{*}{SNR [dB]} &
	\\
\cline{3-6} \cline{8-11}
	&
	&
	\multicolumn{1}{l}{Train} &
	&
	\multicolumn{2}{c}{Test} &
	&
	\multicolumn{1}{l}{Train} &
	&
	\multicolumn{2}{c}{Test} &
	&
	\\
\cline{3-3} \cline{5-6} \cline{8-8} \cline{10-11}
	&
	&
	\multicolumn{1}{l}{Nrm} &
	&
	\multicolumn{1}{l}{Nrm} &
	\multicolumn{1}{l}{Abn} &
	&
	\multicolumn{1}{l}{Nrm} &
	&
	\multicolumn{1}{l}{Nrm} &
	\multicolumn{1}{l}{Abn} &
	&
	\\
\hline \hline
	\multirow{6}{*}{\rotatebox[origin=c]{90}{Fan}}
	& 00 & $1000$ && $100$ & $100$ & $-9$ & $3$ && $100$ & $100$ & $-9$ & Wind strength variations between domains \\
	& 01 & $1000$ && $100$ & $100$ & $-12$ & $3$ && $100$ & $100$ & $-12$ & Two products from the same manufacturer with size variations between domains \\
	& 02 & $1000$ && $100$ & $100$ & $-9$ & $3$ && $100$ & $100$ & $-9$ & Factory noise variations between domains \\
	& 03 & $1000$ && \multicolumn{2}{c}{200} & $-9 \pm 1.5$ & $3$ && \multicolumn{2}{c}{200} & $-9 \pm 1.5$ & Wind strength variations between domains \\
	& 04 & $1000$ && \multicolumn{2}{c}{200} & $-9 \pm 1.5$ & $3$ && \multicolumn{2}{c}{200} & $-9 \pm 1.5$ & Wind strength variations between domains \\
	& 05 & $1000$ && \multicolumn{2}{c}{200} & $-9 \pm 1.5$ & $3$ && \multicolumn{2}{c}{200} & $-9 \pm 1.5$ & Wind temperature variations between domains \\
\hline
	\multirow{6}{*}{\rotatebox[origin=c]{90}{Gearbox}}
	& 00 & $1001$ && $165$ & $105$ & $-15$ & $3$ && $108$ & $108$ & $-15$ & Voltage variations between domains \\
	& 01 & $1008$ && $108$ & $108$ & $-15$ & $3$ && $108$ & $108$ & $-15$ & Arm-length variations between domains \\
	& 02 & $1008$ && $108$ & $138$ & $-15$ & $3$ && $105$ & $120$ & $-15$ & Weight variations between domains \\
	& 03 & $1005$ && \multicolumn{2}{c}{234} & $-15 \pm 1.5$ & $3$ && \multicolumn{2}{c}{286} & $-15 \pm 1.5$ & Voltage variations between domains \\
	& 04 & $1092$ && \multicolumn{2}{c}{252} & $-15 \pm 1.5$ & $3$ && \multicolumn{2}{c}{234} & $-15 \pm 1.5$ & Arm-length variations between domains \\
	& 05 & $1008$ && \multicolumn{2}{c}{210} & $-17 \pm 1.5$ & $3$ && \multicolumn{2}{c}{212} & $-17 \pm 1.5$ & Voltage and arm-length variations between domains \\
\hline
	\multirow{6}{*}{\rotatebox[origin=c]{90}{Pump}}
	& 00 & $1000$ && $100$ & $100$ & $-9$ & $3$ && $100$ & $100$ & $-9$ & Submersible pump; viscosity variations b/w domains \\
	& 01 & $1000$ && $100$ & $100$ & $-12$ & $3$ && $100$ & $100$ & $-18$ & SNR variations between domains \\
	& 02 & $1000$ && $100$ & $100$ & $-9$ & $3$ && $100$ & $100$ & $-9$ & Multiple pumps running simultaneously in the target domain; anomaly condition indicating an abnormality in one or more of the pumps \\
	& 03 & $1000$ && \multicolumn{2}{c}{200} & $-11 \pm 1.5$ & $3$ && \multicolumn{2}{c}{200} & $-11 \pm 1.5$ & Submersible pump; viscosity variations b/w domains \\
	& 04 & $1000$ && \multicolumn{2}{c}{200} & $-9 \pm 1.5$ & $3$ && \multicolumn{2}{c}{200} & $-9 \pm 1.5$ & Factory noise variations between domains \\
	& 05 & $1000$ && \multicolumn{2}{c}{200} & $-9 \pm 1.5$ & $3$ && \multicolumn{2}{c}{200} & $-9 \pm 1.5$ & Multiple pumps running simultaneously in the target domain; anomaly condition indicating an abnormality in one or more of the pumps \\
\hline
	\multirow{6}{*}{\rotatebox[origin=c]{90}{Slide rail}}
	& 00 & $1000$ && $100$ & $100$ & $-12$ & $3$ && $100$ & $100$ & $-12$ & Ball screw type; velocity variations between domains \\
	& 01 & $1000$ && $100$ & $100$ & $-12$ & $3$ && $98$ & $100$ & $-12$ & Ball screw type; operation mode changes b/w domains \\
	& 02 & $1000$ && $110$ & $102$ & $-10$ & $3$ && $102$ & $102$ & $-10$ & Belt type; belt material variations between domains \\
	& 03 & $1000$ && \multicolumn{2}{c}{200} & $-12 \pm 1.5$ & $3$ && \multicolumn{2}{c}{200} & $-12 \pm 1.5$ & Ball screw type; velocity variations between domains \\
	& 04 & $1000$ && \multicolumn{2}{c}{200} & $-14 \pm 1.5$ & $3$ && \multicolumn{2}{c}{200} & $-14 \pm 1.5$ & Ball screw type; operation mode changes b/w domains \\
	& 05 & $1000$ && \multicolumn{2}{c}{204} & $-12 \pm 1.5$ & $3$ && \multicolumn{2}{c}{203} & $-12 \pm 1.5$ & Belt type; belt material variations between domains \\
\hline
	\multirow{6}{*}{\rotatebox[origin=c]{90}{Valve}}
	& 00 & $1000$ && $100$ & $100$ & $-9$ & $3$ && $100$ & $100$ & $-12$ & Open/close timing variations between domains \\
	& 01 & $1000$ && $100$ & $100$ & $-9$ & $3$ && $100$ & $100$ & $-12$ & No pump running in the source domain \\
	& 02 & $1000$ && $100$ & $100$ & $-12$ & $3$ && $100$ & $100$ & $-10$ & Two valves running simultaneously in the target domain; anomaly condition indicates a small piece of metal is caught in one of the valves \\
	& 03 & $1000$ && \multicolumn{2}{c}{200} & $-9 \pm 1.5$ & $3$ && \multicolumn{2}{c}{200} & $-9 \pm 1.5$ & Open/close timing variations between domains \\
	& 04 & $1000$ && \multicolumn{2}{c}{200} & $-9 \pm 1.5$ & $3$ && \multicolumn{2}{c}{200} & $-9 \pm 1.5$ & No pump running in the source domain; water flowing in the target domain \\
	& 05 & $1000$ && \multicolumn{2}{c}{200} & $-12 \pm 1.5$ & $3$ && \multicolumn{2}{c}{200} & $-12 \pm 1.5$ & Two valves running simultaneously in the target domain; anomaly condition indicates a small piece of metal is caught in one of the valves \\
\hline
	\multicolumn{2}{l ||}{Total} & $30122$ && \multicolumn{2}{c}{6274} & \multicolumn{1}{c |}{---} & $90$ && \multicolumn{2}{c}{6182} & \multicolumn{1}{c |}{---} & \multicolumn{1}{c}{---} \\
\hline
\end{tabular}
\end{center}
\end{table*}

\section{MIMII DUE Dataset} 
\label{sec:dataset}

\subsection{Overview of Dataset} 
\label{sec:overview}

\begin{figure}[t]
\begin{center}
\begin{minipage}{0.49\hsize}
\begin{center}
\includegraphics[width=0.97\hsize,clip]{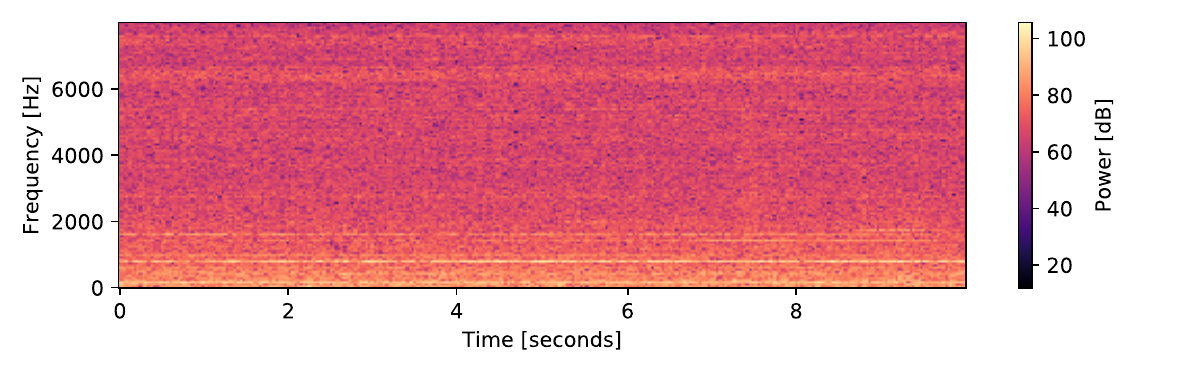}\\
(a) Fan\\
\end{center}
\end{minipage}
\begin{minipage}{0.49\hsize}
\begin{center}
\includegraphics[width=0.97\hsize,clip]{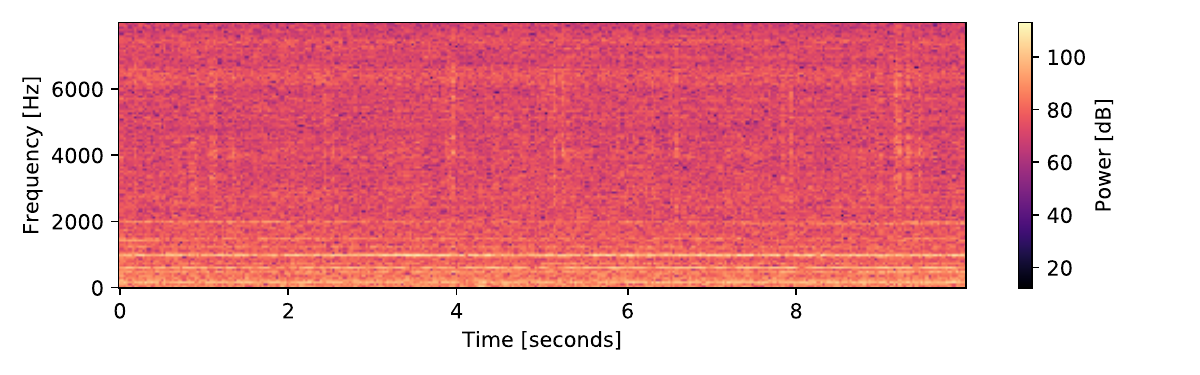}\\
(b) Gearbox\\
\end{center}
\end{minipage}
\begin{minipage}{0.49\hsize}
\begin{center}
\includegraphics[width=0.97\hsize,clip]{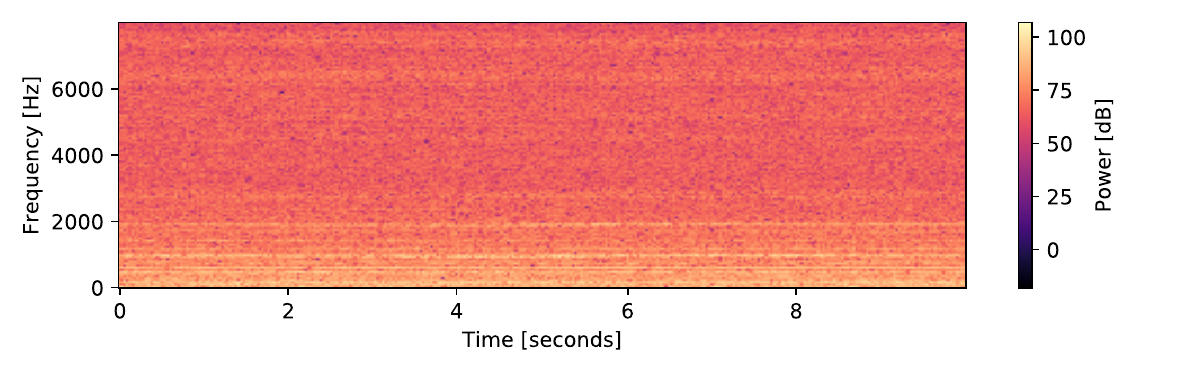}\\
(c) Pump\\
\end{center}
\end{minipage}
\begin{minipage}{0.49\hsize}
\begin{center}
\includegraphics[width=0.97\hsize,clip]{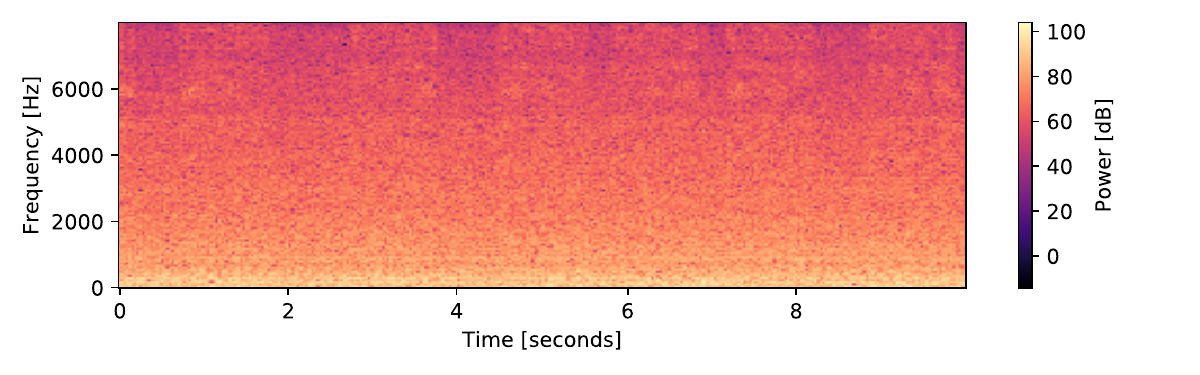}\\
(d) Slide rail\\
\end{center}
\end{minipage}
\begin{minipage}{0.49\hsize}
\begin{center}
\includegraphics[width=0.97\hsize,clip]{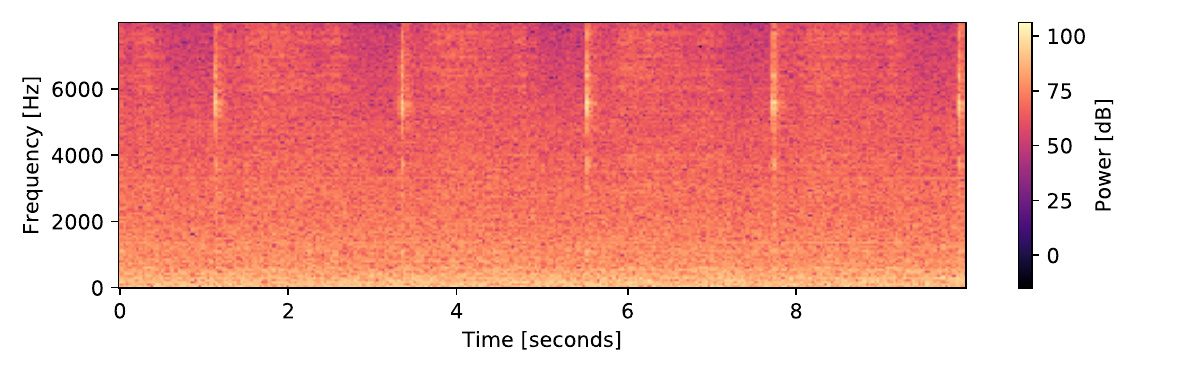}\\
(e) Valve\\
\end{center}
\end{minipage}
\caption{Power spectrograms under normal conditions}
\label{fig:spec}
\end{center}
\end{figure}

Table \ref{tab:overview} summarizes the content of the MIMII DUE dataset.
The dataset consists of normal and abnormal operating sounds of five different types of industrial machines: fans, gearboxes, pumps, slide rails, and valves.
The description of each machine type is as follows.
\begin{description}[style=unboxed,leftmargin=0cm]
\item[Fan] An industrial fan used to keep gas or air flowing in a factory. The strength and temperature of the wind can be changed.
\item [Gearbox] A gearbox that links a direct current (DC) motor to a slider-crank mechanism, transmitting the power generated by the rotation of the motor at a constant speed to the slider-crank mechanism.
The slider-crank mechanism then converts the rotational motion into a linear motion and raises and lowers its weight.
\item [Pump] A submersible or land-based water pump that continuously takes in and discharges water from a pool.
\item [Slide rail] A linear slide system consisting of a moving platform and a staging base that repeats a pre-programmed operation pattern.
\item [Valve] A solenoid valve that repeatedly opens and closes according to a pre-programmed operating pattern and is connected to a pump to control air or water flow.
\end{description}
The sounds emitted by these machines may be stationary or non-stationary, have different features, and have different degrees of difficulty when it comes to identifying them.
Figure \ref{fig:spec} depicts a power spectrogram of the sounds of all machine types, clearly showing that each machine has its unique sound characteristics.

As shown in Table \ref{tab:overview}, the data for each machine type include six subsets, 00 through 05, called ``sections'', and each section roughly corresponds to a single product.
The only exception is ``fan 01'', which contains two products from the same manufacturer.
Each section consists of data from two domains, i.e., the source domain and the target domain.
The source domain means the original condition with enough training clips, and the target domain means the changed condition where only a few audio clips are available as training data. 
This data imbalance reflects the fact that, in the real world, sufficient training data for the target domain are often not available.
The conditions of the source and target domains differ in terms of operating speed, machine load, viscosity, heating temperature, environmental noise, SNR, etc.

The entire dataset contains more than 42,000 clips, each containing 10 seconds of single-channel 16-bit audio data sampled at 16 kHz.
As shown in Table \ref{tab:overview}, the training data for the source domain in any section consists of a minimum of 1000 clips. 
The training data for the target domain, on the other hand, contain only three clips in any section. 
Since the dataset is intended for use in evaluating unsupervised ASD, all training data are normal sounds.
The test data for each domain contain around 100 clips each for both the normal and real anomalous conditions.
Table \ref{tab:anomaly} lists the anomalous conditions included in the dataset.

\setlength{\tabcolsep}{1mm} 
\begin{table}[t]
\centering
\caption{Anomalous conditions for each machine type}
\label{tab:anomaly}
\footnotesize
\begin{tabular}{l c}
\hline
\begin{tabular}{@{\hskip0pt}l@{\hskip0pt}} Machine \\ type \end{tabular} & 
\begin{tabular}{c} Anomalous conditions \end{tabular} \\
\hline
Fan & 
\begin{tabular}{c} Wing damage, unbalanced, clogging, and over voltage \end{tabular} \\
Gearbox & 
\begin{tabular}{c} Gear damage, overload, over voltage, etc. \end{tabular} \\
Pump & 
\begin{tabular}{c} Contamination, clogging, leakage, dry run, etc. \end{tabular} \\
Slide rail & 
\begin{tabular}{c} Rail damage, loose belt, no grease, etc. \end{tabular} \\
Valve & 
\begin{tabular}{c} Contamination \end{tabular} \\
\hline
\end{tabular}
\end{table}

In the training data, attribute information for each clip (other than normal or abnormal) is indicated by its file name. 
We have made the attribute information public considering its potential use in future research on representation learning for machine sounds.

\subsection{Recording Setup and Post-Processing Procedures} \label{sec:rec}

We describe the recording and post-processing procedures here.
As mentioned above, we recorded 16-bit audio clips at 16 kHz. 
We used a TAMAGO-03 microphone array, manufactured by System In Frontier Inc.~\cite{tamago}. 
However, this dataset contains only the single-channel audio recorded with the first microphone of the array. 
Table \ref{tab:rec} shows the differences in recording settings for each machine type.
The gearboxes and slide rails were recorded in an anechoic chamber, and we convolved an impulse response with the recorded sounds.
The impulse response was generated by Pyroomacoustics~\cite{scheibler2018pyroomacoustics} such that the RT60 is 0.3 seconds.
The other machines can not run in an anechoic chamber, so they were recorded in ordinary rooms or a sound isolation booth, and we did not convolve an impulse response with their sounds.

\setlength{\tabcolsep}{1mm} 
\begin{table}[t]
\centering
\caption{Differences in recording settings for each machine type}
\label{tab:rec}
\footnotesize
\begin{tabular}{lcc} \hline
\begin{tabular}{@{\hskip0pt}l@{\hskip0pt}} Machine \\ type \end{tabular} &
Room &
\begin{tabular}{c} Distance from \\ microphone [m] \end{tabular}  \\ \hline
Fan & Sound isolation booth & 0.6 \\
Gearbox & Anechoic chamber & 0.6 \\
Pump & \begin{tabular}{c} Ordinary room with reverberation \end{tabular} & 0.6 -- 1.0 \\
Slide rail & Anechoic chamber & 0.6 \\
Valve & \begin{tabular}{c} Ordinary room with reverberation \end{tabular} & 0.6 \\ \hline
\end{tabular}
\end{table}

Apart from the machine sounds, background noise in multiple real factories was recorded and later mixed with the machine sounds to simulate real environments.
The same microphone used to record the machine sounds was also used to record the background noise.
The noise-mixed data of each section are made by the following steps.
{
	\setlength{\leftmargini}{12pt}
	\begin{enumerate}
	\item The average power over all clips in the section, $a$, was calculated.
	\item For each clip $i$ from the section,
	{
		\begin{enumerate}
		\setlength{\leftskip}{-0.3cm}
		\item the signal-to-noise ratio (SNR) $\gamma$ dB of the clip was determined according to the range of SNR for each section defined as shown in Table \ref{tab:overview},
		\item a background-noise clip $j$ was randomly selected, and its power $b_j$ was tuned so that $\gamma = 10 \log_{10}\left( a /b_j \right)$, and
		\item the noise-mixed data was generated by mixing the machine-sound clip $i$ and the power-tuned background-noise clip $j$.\
		\end{enumerate}
	}
	 \end{enumerate}
}
To make domain shifts in background noise, as shown in Table \ref{tab:overview}, 
we made sure that the factories where we recorded the noise used in the target domains of ``fan 02'' and ``pump 04'' were different from the factories where we recorded the noise used in the other data.
Also, we changed the SNR between domains in ``pump 01''.

\section{Experiments}\label{sec-eval}

To enable usage of the dataset, we have made two simple baseline systems available at \url{https://github.com/y-kawagu/dcase2021_task2_baseline_ae} and \url{https://github.com/y-kawagu/dcase2021_task2_baseline_mobile_net_v2}, 
and they are also the baseline systems for Task 2 of the DCASE 2021 Challenge. 
Both baseline systems consist of Python codes for training and testing. 

The first baseline system is an autoencoder-based ASD. 
The frame size for a short-time Fourier transform (STFT) is 64 ms and the hop size is 32 ms. 
The input vector of the autoencoder is five consecutive time-frames of a 128-bin log-mel-spectrogram.
The autoencoder model consists of an input fully connected (FC) layer, four 128-unit FC layers, an 8-unit bottleneck layer, four 128-unit FC layers, and an output 640-unit FC layer.
The rectified linear unit (ReLU)~\cite{nair2010relu} is installed after each FC layer except the output layer.
The batch size is 512, the Adam optimizer~\cite{kingma2015adam} is used, the learning rate is 0.001, and the training process stops after 100 epochs.
The anomaly score is calculated as the averaged reconstruction error.

The other baseline system is a machine identification-based ASD using MobileNetV2~\cite{sandler2018mobilenetv2}.
Many teams have had success with similar approaches in DCASE 2020 Challenge Task 2~\cite{giri2020self, primus2020anomalous, inoue2020detection}.
The frame size for an STFT is 64 ms and the hop size is 32 ms. 
The input image of the MobileNetV2 is 64 consecutive time-frames of a 128-bin log-mel-spectrogram.
The MobileNetV2 is trained such that it identifies from which section the sound was generated.
The batch size is 32, the Adam optimizer is used, the learning rate is 0.00001, and the training process stops after 20 epochs.
The anomaly score is calculated as the average negative logit of the predicted probabilities for the correct section.

Table \ref{tab:result} shows the area under the curve (AUC) scores. 
We trained one model for sections 00--02 of each machine type for both baseline systems and calculated the AUC for each section
\footnote{Sections 03--05 are a subset of the evaluation dataset in DCASE 2021 Challenge Task 2, so only the results for sections 00--02 are shown until the end of the challenge.}.
As shown in Table \ref{tab:result}, there were  significant gaps in the AUC score between the source and target domains in many sections, 
which indicates that the dataset contains the domain shifts.
In some sections, the AUC of the target domain was higher than the AUC of the source domain, 
but this could be because the target domain in that section happened to be similar to the source domain in other sections.

\begin{table}[t!]
\begin{center}
\caption{AUC scores for baseline methods.}
\label{tab:result}
\footnotesize
\begin{tabular}{l l r r p{2pt} r r}
\hline
\multicolumn{2}{@{}l}{\multirow{2}{*}{ \begin{tabular}{@{\hskip0pt}l@{\hskip0pt}} Machine type/ \\ section index \end{tabular} }} &
\multicolumn{2}{c}{Autoencoder} &&
\multicolumn{2}{c}{MobileNetV2} \\
\cline{3-4} \cline{6-7}
& &
\multicolumn{1}{c}{Source} &
\multicolumn{1}{c}{Target} &&
\multicolumn{1}{c}{Source} &
\multicolumn{1}{c}{Target} \\
\hline
  \multirow{3}{*}{Fan}
  & 00 & $0.667$ & $0.697$ && $0.436$ & $0.533$ \\
  & 01 & $0.674$ & $0.500$ && $0.783$ & $0.781$ \\
  & 02 & $0.642$ & $0.662$ && $0.742$ & $0.604$ \\ \hline
  \multirow{3}{*}{Gearbox}
  & 00 & $0.560$ & $0.743$ && $0.814$ & $0.750$ \\
  & 01 & $0.728$ & $0.721$ && $0.607$ & $0.563$ \\
  & 02 & $0.590$ & $0.664$ && $0.716$ & $0.644$ \\ \hline
  \multirow{3}{*}{Pump}
  & 00 & $0.675$ & $0.580$ && $0.641$ & $0.591$ \\
  & 01 & $0.824$ & $0.474$ && $0.863$ & $0.719$ \\
  & 02 & $0.639$ & $0.628$ && $0.537$ & $0.502$ \\ \hline
  \multirow{3}{*}{Slide rail}
  & 00 & $0.741$ & $0.672$ && $0.615$ & $0.520$ \\
  & 01 & $0.822$ & $0.669$ && $0.800$ & $0.468$ \\
  & 02 & $0.783$ & $0.462$ && $0.799$ & $0.556$ \\ \hline
  \multirow{3}{*}{Valve}
  & 00 & $0.503$ & $0.471$ && $0.583$ & $0.522$\\
  & 01 & $0.535$ & $0.564$ && $0.536$ & $0.686$\\
  & 02 & $0.599$ & $0.552$ && $0.561$ & $0.536$\\ \hline
\end{tabular}\\
\end{center}
\end{table}

\begin{figure}[t!]
\begin{center}
\begin{minipage}{0.88\hsize}
\begin{center}
\includegraphics[width=0.97\hsize,clip]{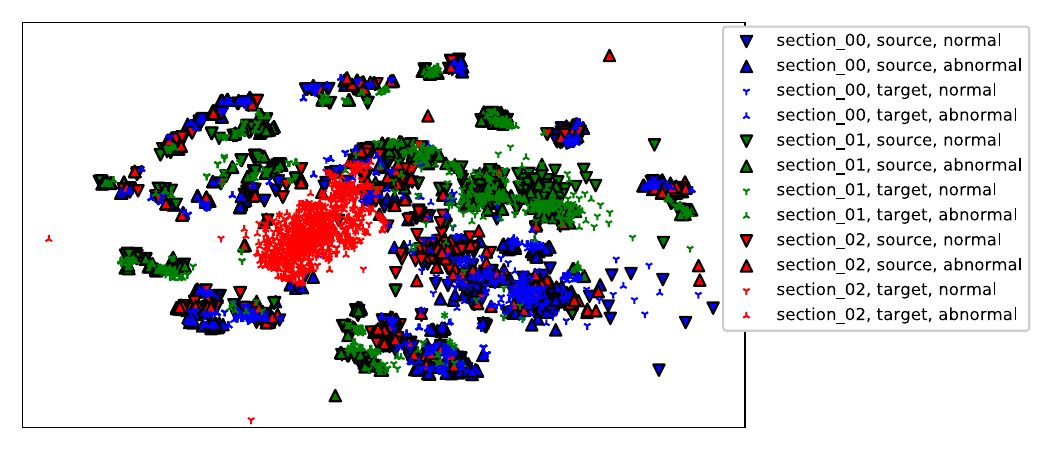}\\
(a) Fan\\
\end{center}
\end{minipage}
\begin{minipage}{0.88\hsize}
\begin{center}
\includegraphics[width=0.97\hsize,clip]{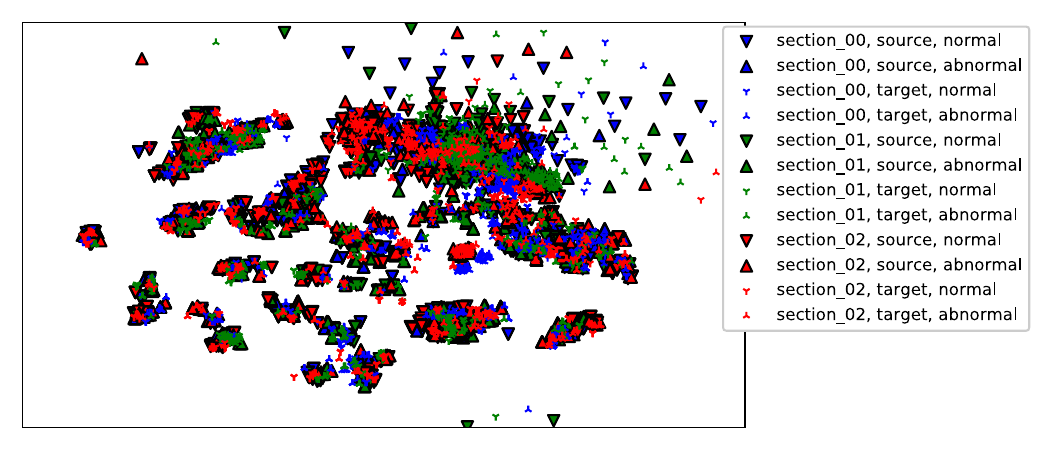}\\
(b) Gearbox\\
\end{center}
\end{minipage}
\begin{minipage}{0.88\hsize}
\begin{center}
\includegraphics[width=0.97\hsize,clip]{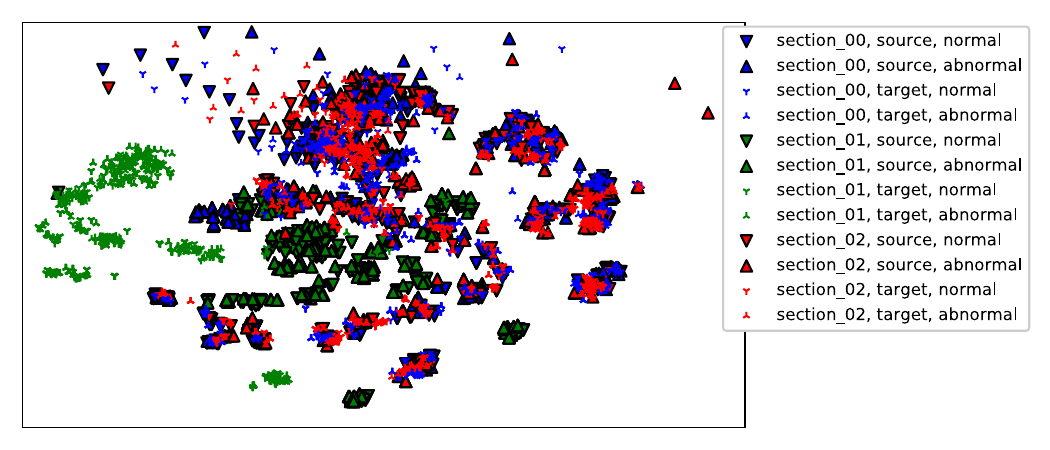}\\
(c) Pump\\
\end{center}
\end{minipage}
\begin{minipage}{0.88\hsize}
\begin{center}
\includegraphics[width=0.97\hsize,clip]{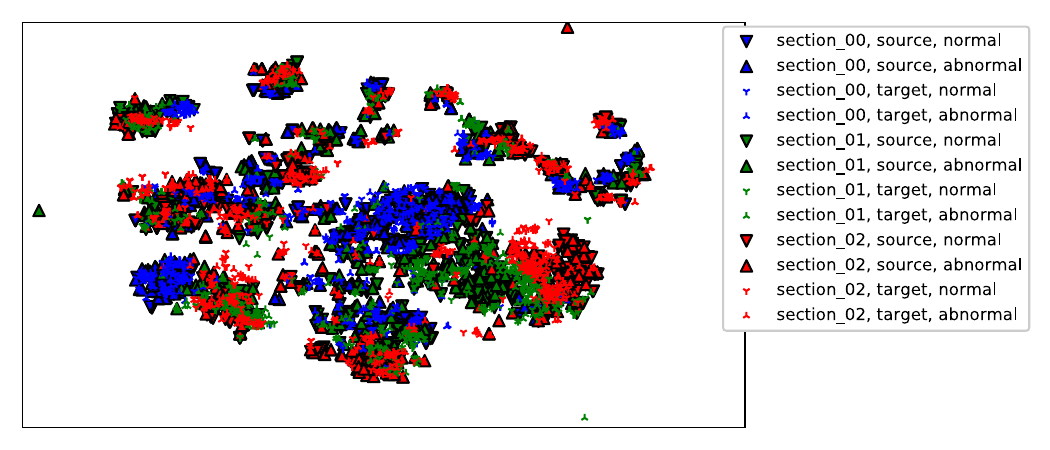}\\
(d) Slide rail\\
\end{center}
\end{minipage}
\begin{minipage}{0.88\hsize}
\begin{center}
\includegraphics[width=0.97\hsize,clip]{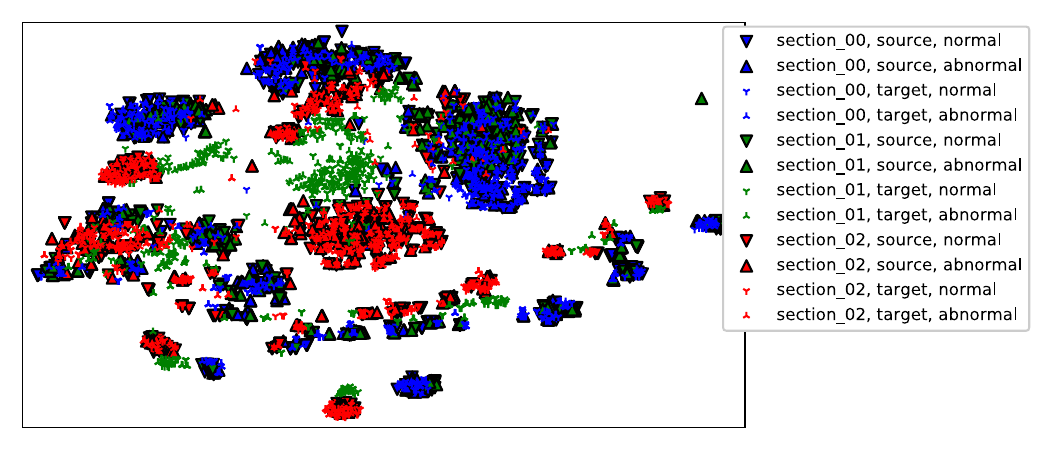}\\
(e) Valve\\
\end{center}
\end{minipage}
\caption{Two-dimensional scatter plot based on t-SNE. X and Y-axes show the first and second dimensions of t-SNE, respectively.}
\label{fig:tsne}
\end{center}
\end{figure}

Figure \ref{fig:tsne} shows two-dimensional scatter plots of test data in sections 00--02 using t-distributed stochastic neighbor embedding (t-SNE) to check the change in the distribution of the acoustic features due to domain shifts.
The input vectors for t-SNE are the same as the MobileNetV2-based baseline.
As we can see, normal and abnormal samples belong to the same cluster in many machine types and sections, while source and target domains belong to different clusters.
These results indicate that the dataset contains domain shifts, which can be helpful for checking the robustness against domain shifts.

\section{Conclusion}

We introduced a new dataset called ``MIMII DUE'' designed for checking the robustness of anomalous sound detection against domain shifts.
The dataset consists of normal and abnormal operating sounds of five different types of industrial machines, with domain shifts occurring.
The experimental results indicate that the dataset contains domain shifts, which can be helpful for checking the robustness against domain shifts.

\bibliographystyle{IEEEtran}
\bibliography{refs}

\end{sloppy}
\end{document}